# Phenomenology of Hidden Abelian Higgs Model at Hadron Colliders using Neural Networks and Monte Carlo Techniques
## (Gauge Sector $Z_{HAHM}$ & Scalar Sector H)

Nady Bakhet[1,2], Maxim Yu Khlopov[3,4], Tarek Hussein[1,2]

[1]Department of Physics, Cairo University, Giza, Egypt
[2]Egyptian Network of High Energy Physics - ASRT, Cairo, Egypt
[3]APC Laboratory, IN2P3/CNRS, Paris, France
[4]National Research Nuclear University "MEPHI" (Moscow Engineering Physics Institute) Russia

**ABSTRACT**

The Hidden Abelian Higgs Model (HAHM) is an extension of the Standard Model based on the gauge group G = $SU(3)_C \otimes SU(2)_L \otimes U(1)_Y \otimes U(1)_X$ i.e. an extra $U(1)_X$ gauge group in addition to the SM gauge groups. In HAHM, an Abelian Hidden sector is coupled to the SM, and the resulting new Higgs boson H and new neutral gauge boson $Z_{HAHM}$ fields are allowed to mix with the corresponding SM fields. In this work, we search for the new neutral gauge boson $Z_{HAHM}$ and new Higgs boson in the context of the Hidden Abelian Higgs Model (HAHM) through the decay process $pp(\bar{p}) \rightarrow H \rightarrow Z_{HAHM}Z_{HAHM} \rightarrow \ell^\pm\ell^\mp\ell^\pm\ell^\mp, (\ell = e, \mu)$ with **4 charged leptons** in the final state in the mass range $200\ GeV \leq Z_{HAHM} \leq 700\ GeV$ and the production and decay of new Higgs boson with **2 charged leptons** in the final state of the process $pp(\bar{p}) \rightarrow H \rightarrow \ell^\pm\ell^\mp jj$, $(\ell = e, \mu)$ with jets or final state of the process $pp(\bar{p}) \rightarrow H \rightarrow \ell^\pm\ell^\mp \nu\bar{\nu}$, $(\ell = e, \mu)$ with missing energy. We used Monte Carlo data produced from proton-proton collisions at the LHC CERN (CMS and ATLAS detectors) for different center of mass energies up to $\sqrt{s} = 14\ TeV$ and also Monte Carlo data produced from proton-antiproton collisions at the Tevatron Fermi Lab (CDF and D0 detectors) at $\sqrt{s}$ =1.96 TeV. To discriminate the signal events comes from the new gauge boson $Z_{HAHM}$ and new Higgs boson than the SM background events, we used the Artificial Neural Networks (ANNs). The distributions of the different kinematics, invariant mass, transverse momentum, transverse energy and pseudorapidity of the electrons and muons in the final states $\ell^\pm\ell^\mp\ell^\pm\ell^\mp$, $\ell^\pm\ell^\mp jj$ and $\ell^\pm\ell^\mp \nu\bar{\nu}$ where $\ell = e, \mu$ are presented.

# 1 Hidden Abelian Higgs Model (HAHM)

The Standard Model of particle physics [1–3] has been tested by different experiments as the Tevatron at FermiLab and the LHC at CERN. Also the Standard Model of particles physics cannot give explanations for some problems as the operation of symmetry breaking, mass generation, the hierarchy problem, the flavor problem, the baryon asymmetry problem and the dark matter problem. The LHC at CERN and the Tevatron at Fermi Lab may discover an unexpected and a solution to one of these problems and expect new particles that may solve the problems. The Hidden Abelian Higgs Model (HAHM) [4, 5] is an extension of the Standard Model based on the gauge group G = $SU(3)_C \otimes SU(2)_L \otimes U(1)_Y \otimes U(1)_X$ where an extra $U(1)_X$ group in addition to the SM symmetry gauge groups. The SM fields are singlet under the $U(1)_X$ group and the SM fields couple to the new gauge sector via mixing effects in the gauge and scalar sectors. In HAHM, an Abelian Hidden sector is coupled to the Standard Model and will result a new scalar filed (Higgs boson) and new neutral gauge boson $Z_{HAHM}$ fields and they are mix with the corresponding SM fields. The Hidden sector which connected to the Standard Model, where the Higgs domains of the two sectors are mixing as in different beyond the SM models [6, 7], and it has been studied in [8, 9]. The word "hidden" of HAHM means that there are a new particles (new Higgs and new gauge boson) rather than the SM particles and these new particles do not have a charge under the SM gauge groups G = $SU(3)_C \otimes SU(2)_L \otimes U(1)_Y$ and these new particles do not couple to the Standard Model particles via gauge interactions. The lagrangian of the Hidden Abelian Higgs Model may help to find solutions for the mentioned problems of the SM at both the LHC and Tevatron. In HAHM, the new neutral gauge boson $Z_{HAHM}$ that couples to SM states according to the strength of the kinetic mixing parameter and the new two CP-even Higgs boson mass eigenstates couple to the Standard Model states by the mixing of the HAHM Higgs boson with the SM Higgs boson. The gauge group of The Hidden Abelian Higgs Model (HAHM) [4, 5] is:

$$G = SU(3)_C \otimes SU(2)_L \otimes U(1)_Y \otimes U(1)_X \quad (1)$$

In HAHM, The new neutral gauge $Z_{HAHM}$ couples to the SM is via kinetic mixing parameter with the hypercharge gauge boson $B_\mu$ and the new Higgs boson $\Phi_H$ in addition to the usual SM Higgs boson $\Phi_{SM}$ couple through the term $|\Phi_{SM}|^2|\Phi_H|^2$. The mass mixing between the SM Higgs $\Phi_{SM}$ and $\Phi_H$, which results in two mass eigenstates, h, H. and the two real physical Higgs boson $\Phi_{SM} \approx \Phi_H$ mix after symmetry breaking.

The mass eigenstates h and H are:

$$\begin{pmatrix} \Phi_{SM} \\ \Phi_H \end{pmatrix} = \begin{pmatrix} c_h & s_h \\ -s_h & c_h \end{pmatrix} \begin{pmatrix} h \\ H \end{pmatrix} \quad (2)$$

The neutral gauge boson mass matrix is:

$$\begin{pmatrix} B \\ W^3 \\ X \end{pmatrix} = \begin{pmatrix} c_W & -s_W c_\alpha & s_W c_\alpha \\ s_W & s_W c_\alpha & -c_W s_\alpha \\ 0 & s_\alpha & c_\alpha \end{pmatrix} \begin{pmatrix} A \\ Z \\ Z_{HAHM} \end{pmatrix} \quad (3)$$

Where $s_x = \sin \theta_x$, $c_x = \cos \theta_x$

In general, the phenomenology of the HAHM at the LHC and the Tevatron is two new physical states, neutral gauge boson $Z_{HAHM}$ and Higgs boson H where $Z_{HAHM}$ is new gauge boson that interacts with the Standard Model fields due to the gauge invariant, kinetic mixing with hypercharge, and H new Higgs boson that interacts with the Standard Model fields due to modulus-squared mixing with the Standard Model Higgs boson.

## 2 Gauge Sector - New Neutral Gauge Boson $Z_{HAHM}$

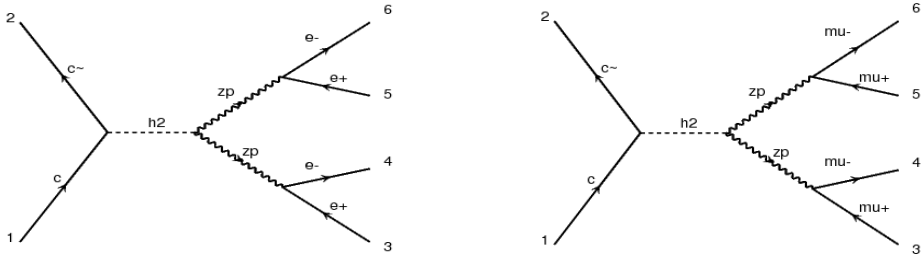

FIG. 1: : Feynman diagrams (Signal) of new gauge boson $Z_p$ pair production and decay to four charged leptons in HAHM via new Higgs boson h2.

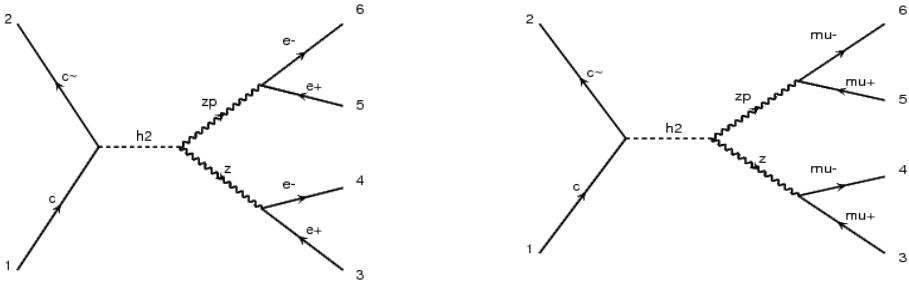

FIG. 2: : Feynman diagrams (Signal) of new gauge boson $Z_p$ production mixing with SM Z gauge boson and decay to four charged leptons in HAHM via new Higgs boson h2.

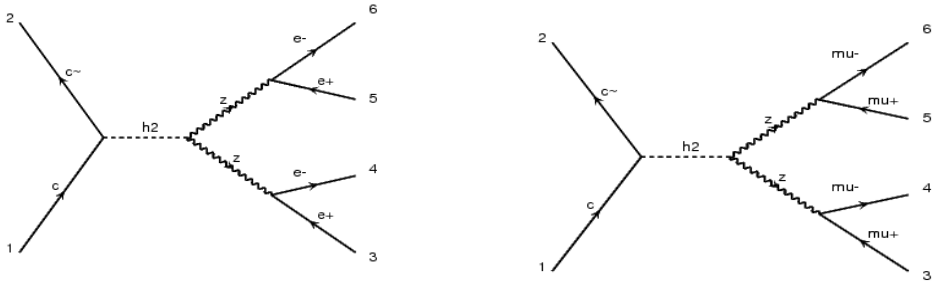

FIG. 3: : Feynman diagrams (Background) of the SM Z gauge boson pair production and decay to four charged leptons in HAHM via new Higgs boson h2.

## 2.1 Production of $Z_{HAHM}$ Boson

In this section, we present an analysis for production of new neutral gauge boson $Z_{HAHM}$ like the SM Z boson in the context of the Hidden Abelian Higgs Model (HAHM) via the decay process $pp(\bar{p}) \to H \to Z_{HAHM} Z_{HAHM} \to \ell^\pm \ell^\mp \ell^\pm \ell^\mp, (\ell = e, \mu)$ in the mass range $200\ GeV \leq Z_{HAHM} \leq 700\ GeV$ using Monte Carlo data produced from proton-proton collisions at the LHC CERN (CMS and ATLAS detectors) for different center of mass energies $\sqrt{s} = 10, 12\ and\ 14\ TeV$ and also Monte Carlo data produced from proton-antiproton collisions at the Tevatron FermiLab (CDF and D0 detectors) at $\sqrt{s} = 1.96$ TeV. Also, to discriminate the signal events comes from the new gauge boson $Z_{HAHM}$ than the SM background events; we used an efficient technique, Artificial Neural Networks (ANNs). In current process of $Z_{HAHM}$ production, The dominant backgrounds come from the VBF process production the SM Z boson and the pair production of $Z_{HAHM}$ boson in this process produce from the decay of the new Higgs boson in HAHM. Figure 4 shows the pair production cross section of new gauge boson $Z_{HAHM}$ in Hidden Abelian Higgs Model (HAHM) at the LHC for $\sqrt{s} = 10, 12$ and 14 TeV and at the Tevatron for $\sqrt{s} = 1.96$ TeV. The values of cross section at the LHC are near for different center of mass energies and far from the values that at the Tevatron. The pair of $Z_{HAHM}$ bosons decay to four charged lepton in final state. Also Figure 5 shows the production cross section of new boson $Z_{HAHM}$ mixing with the SM Z boson in Hidden Abelian Higgs Model (HAHM) at the LHC for $\sqrt{s} = 14$ TeV and at the Tevatron for $\sqrt{s} = 1.96$ TeV with final state of four charged leptons.

The distributions of the invariant mass of the $(\ell^\pm \ell^\mp \ell^\pm \ell^\mp), (\ell = e, \mu)$ of the $Z_{HAHM} Z_{HAHM}$ events and different kinematics of the fours leptons from the process $pp(\bar{p}) \to H \to Z_{HAHM} Z_{HAHM} \to \ell^\pm \ell^\mp \ell^\pm \ell^\mp, (\ell = e, \mu)$ are presented. Our analysis is based on selecting events with 4 charged leptons in the three final states $e^- e^+ e^- e^+, \mu^- \mu^+ \mu^- \mu^+\ and\ e^- e^+ \mu^- \mu^+$ and forming lepton pairs. One pair has a mass m close to one $Z_{HAHM}$ boson mass, and the other has a mass M close to the other $Z_{HAHM}$. The selections rules of events apply on the invariant masses of four leptons are greater than 200 GeV. In the HAHM, the hidden width is known as a function of the three free parameters: $M_H$, $M_{Z\prime}$, and $s_h^2$. The Higgs boson mass is fixed to 300 GeV. The reason is to look specifically at the $Z_{HAHM}$ mass is as in [10, 11]. Figure 7 shows the width of $Z_{HAHM}$ as a function of its mass. We investigate the consequences of the addition of this new process to the search for Higgs boson decays to four charged leptons via a pair gauge bosons $Z_{HAHM}$ decaying to 4 leptons (e or μ) final state and the resulting constraints on the Abelian Hidden sector model parameters. The constraints are drawn in terms of Higgs mixing angle and new gauge boson mass $M_Z$, and $s_h^2$. For the pp(p̄) → H → $Z_{HAHM}Z_{HAHM}$ → 4 leptons search, the Monte Carlo events are selected using four charged leptons(e or μ ). We will consider that the new gauge boson $Z_{HAHM}$ decays only to charged leptons with a 100% branching fraction where Figure 6 shows all decay channels and branching ratios of $Z_{HAHM}$ where the decay channel to quark-antiquark has the highest ratio followed by lepton-antilepton channel. The lightest Higgs boson is mostly the Standard Model and the heavier Higgs boson eigenstate is mostly singlet. The events are classified into two exclusive lepton channels depending on the flavor of the leading lepton, where eμ or μe refers to events with a leading electron or muon. The dilepton invariant mass is required to be greater than 10 GeV. Also there are different analysis sub channels as 4e, 2e2μ and 4μ, arranged by the flavor of the leading lepton pair. The signature for this decay channel is two opposite charge leptons in addition to large transverse momentum and a large momentum imbalance in the event due to the escaping neutrinos. Other backgrounds include Drell-Yan events. Boson pair production (WW and ZZ) can also produce opposite charge lepton pairs with leptons that are not detected. The background rate and composition depend significantly on the jet multiplicity. The two jets selection follows the one jet selection as described with the $p_t^{tot}$ modified to include all selected jets. Several additional criteria are applied to the tag jets defined as the two highest $p_T$ jets in the event. Also these are required to be separated in rapidity by a distance $|\Delta y_{jj}| > 2.5$ and an invariant mass, $M_{jj}$, larger than 500 GeV. On the other hand the leading backgrounds from the Standard Model processes producing two isolated high $p_T$ leptons are WW and the jets are required to have transverse momentum $p_T > 50$ GeV and $|\eta| < 5$ where $\eta = -\ln\tan(\frac{\theta}{2})$ with $\theta$ being the polar angle with respect to the beam. The leptons and photons are required to be separated from jets by $\Delta R > 0.5$ and from one another by $\Delta R > 0.3$ where $\Delta R = \sqrt{(\Delta\eta)^2 + (\Delta\phi)^2}$ and $\phi$ is the azimuthal angle. The jets must be separated from each other by $\Delta R > 0.7$. We need one chaged lepton e or μ with $p_T > 50$ GeV, $|\eta| < 3$ and missing energy transverse $\notin > 50\ GeV$. We need two tagging jets with $|\eta| < 0.2|$. The four charged lepton invariant mass $m_{4\ell} < 200$ GeV and to both lepton pairs for higher masses. The width of the reconstructed mass distribution is dominated by the experimental resolution for $m_H < 350$ GeV and by the natural width of the Higgs boson for higher masses (30 GeV at $m_H = 400$ GeV). The same flavour and oppositecharge lepton pair with an invariant mass closest to the Z boson mass is referred to as the leading lepton pair. The remaining same flavor opposite charge lepton pair is the sub-leading lepton pair. All possible lepton pairs that have the same flavor and opposite charge must satisfy $m_{\ell\ell} > 5\ GeV$ in order to reject.

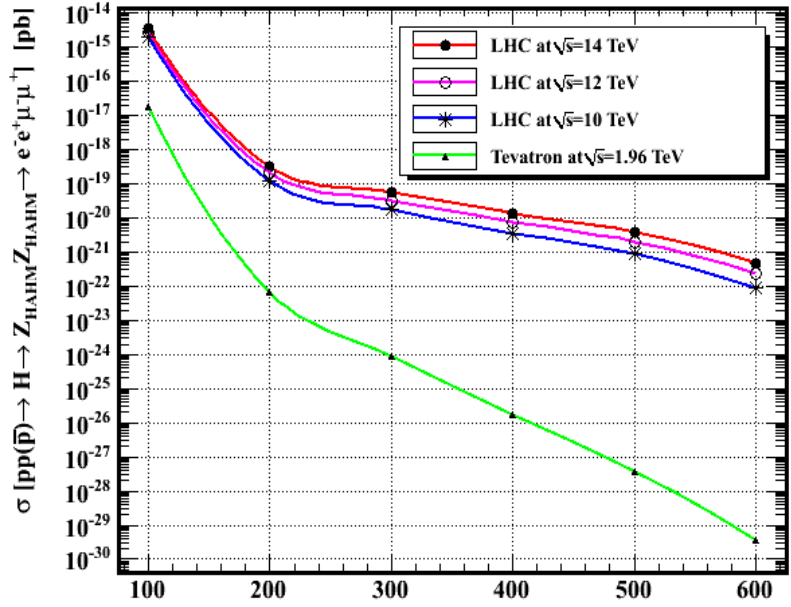

FIG. 4: Pair production cross section of new gauge boson $Z_{HAHM}$ in Hidden Abelian Higgs Model (HAHM) at the LHC for $\sqrt{s} = 10, 12\ and\ 14\ TeV$ and at the Tevatron for $\sqrt{s} = 1.96\ TeV$.

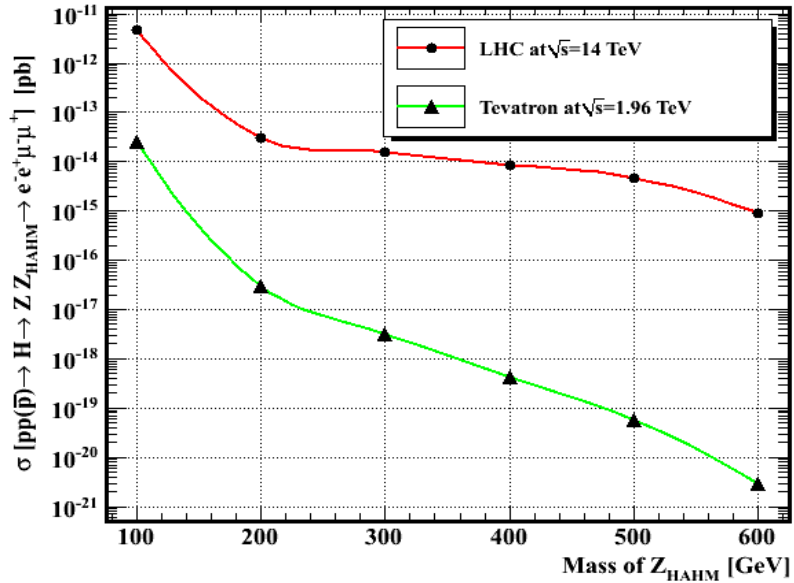

FIG. 5: Production cross section of new boson $Z_{HAHM}$ mixing with Z boson in Hidden Abelian Higgs Model (HAHM) at the LHC for $\sqrt{s} = 14\ TeV$ and at the Tevatron for $\sqrt{s} = 1.96\ TeV$.

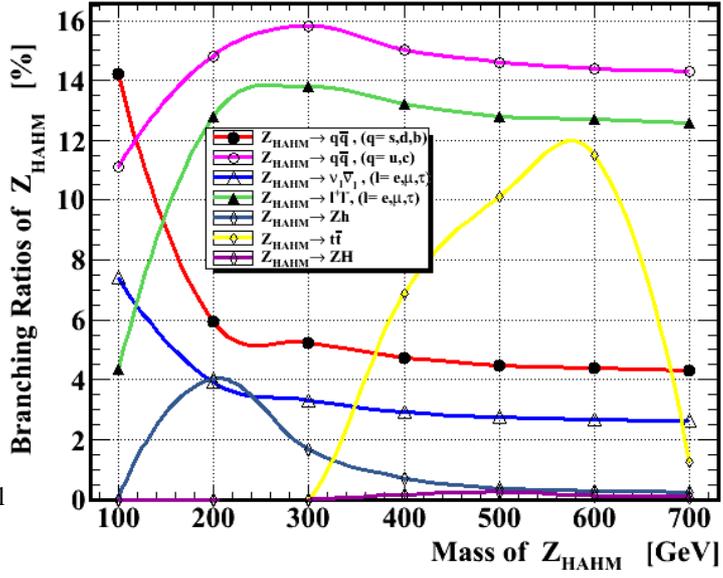

FIG. 6: All branching ratios of $Z_{HAHM}$ boson as a function of the mass.

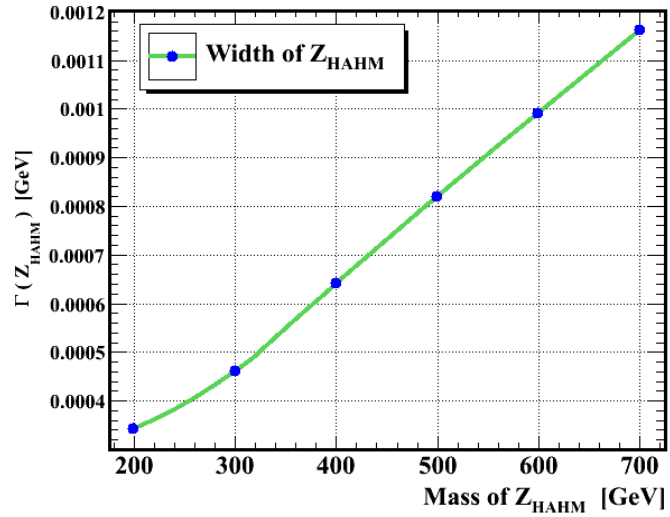

FIG. 7: Decay Width of $Z_{HAHM}$ boson as a function of the mass.

## 2.2 Detection $Z_{HAHM}$ Siganl at Hadron Colliders
### 2.2.1 Process $pp(\bar{p}) \to H \to Z_{HAHM} Z_{HAHM} \to e^-e^+e^-e^+$
**A- Different distributions of ( $e^-e^+e^-e^+$ ) at the LHC (Left) and the Tevatron (Right)**

1-The invariant mass of $e^-e^+e^-e^+$ $final\ state$ **[LHC (Left) & Tevatron (Right)]**

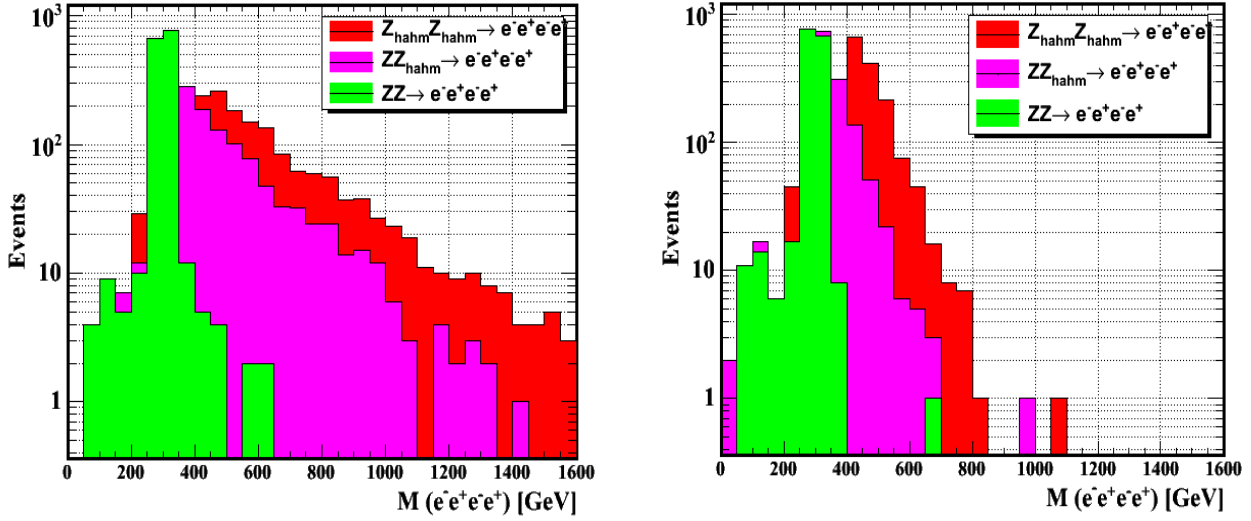

2-Transverse Momentum of $e^-e^+e^-e^+$ $final\ state$ **[LHC (Left) & Tevatron (Right)]**

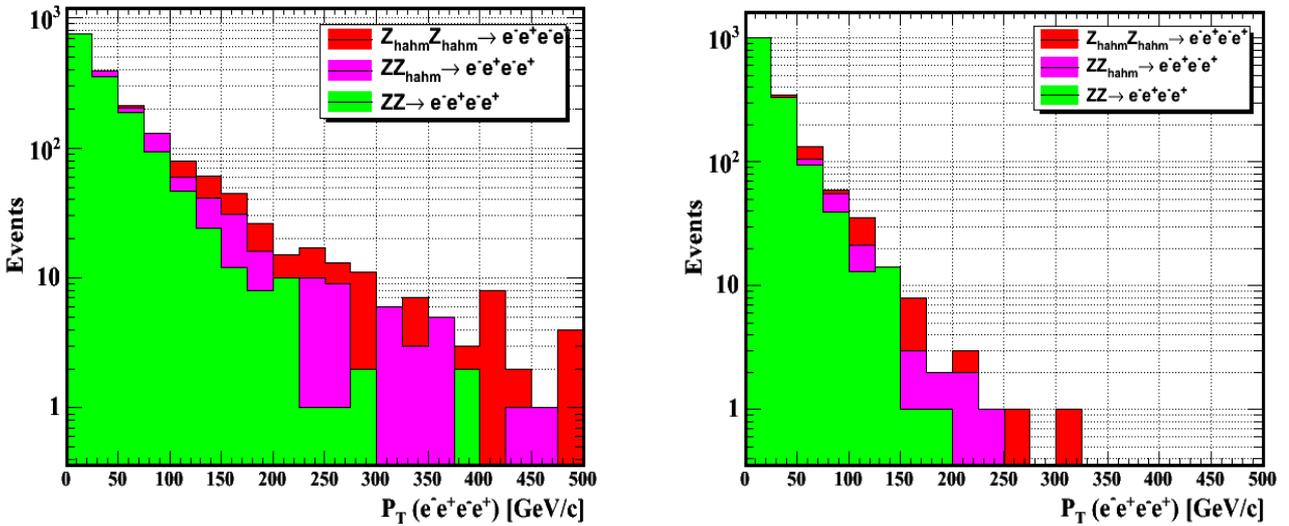

3- Transverse Energy of $e^-e^+e^-e^+$ final state **[LHC (Left) & Tevatron (Right)]**

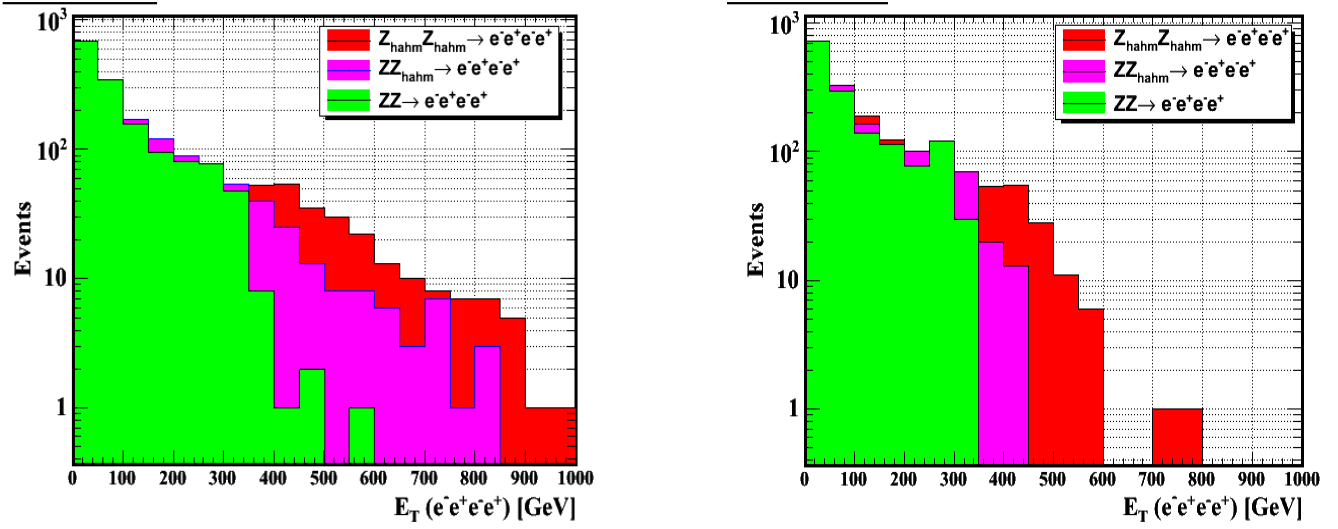

4- Pseudorapidity $e^-e^+e^-e^+\ final\ state$ **[LHC (Left) & Tevatron (Right)]**

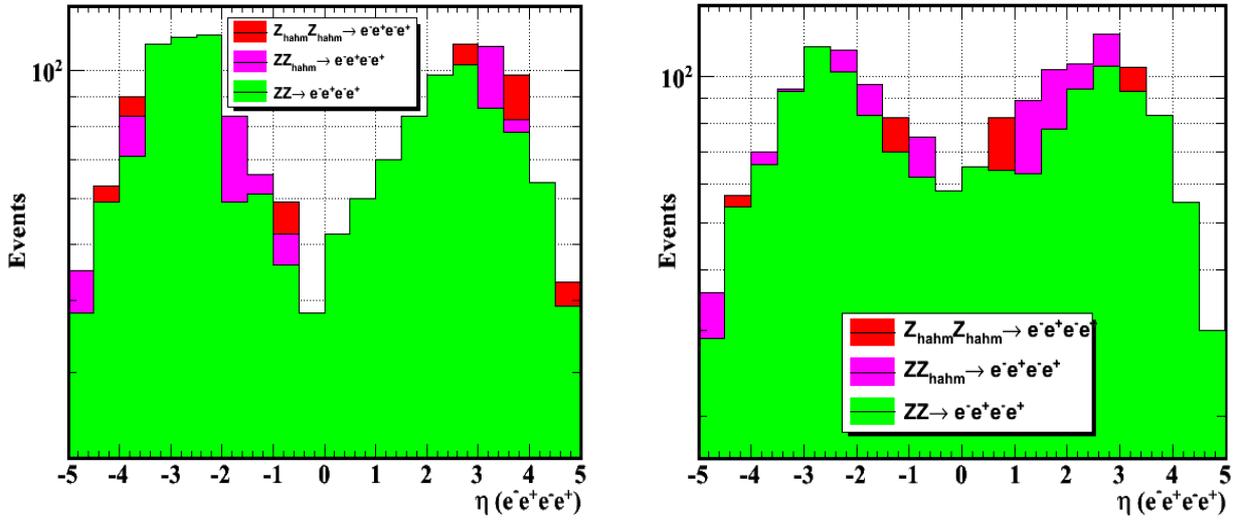

FIG. 8: Comparison between the different distributions of $e^-e^+e^-e^+$ final state produced from the decay of new $Z_{HAHM}$ boson versus background of Z boson in Hidden Abelian Higgs Model (HAHM) at the LHC for $\sqrt{s}=\ 14\ TeV$ (LEFT) and at the Tevatron for $\sqrt{s}=1.96\ TeV$ (RIGHT).

## B-The Discrimination between $Z_{HAHM}$ *Signal*s and *Backgrounds* events using Artificial Neural Networks (ANNs) at the *LHC* (LEFT) and the *Tevatron* (RIGHT)

1- Final discriminations of ANNs Output (signal and background) for $e^-e^+e^-e^+$ $final\ state$

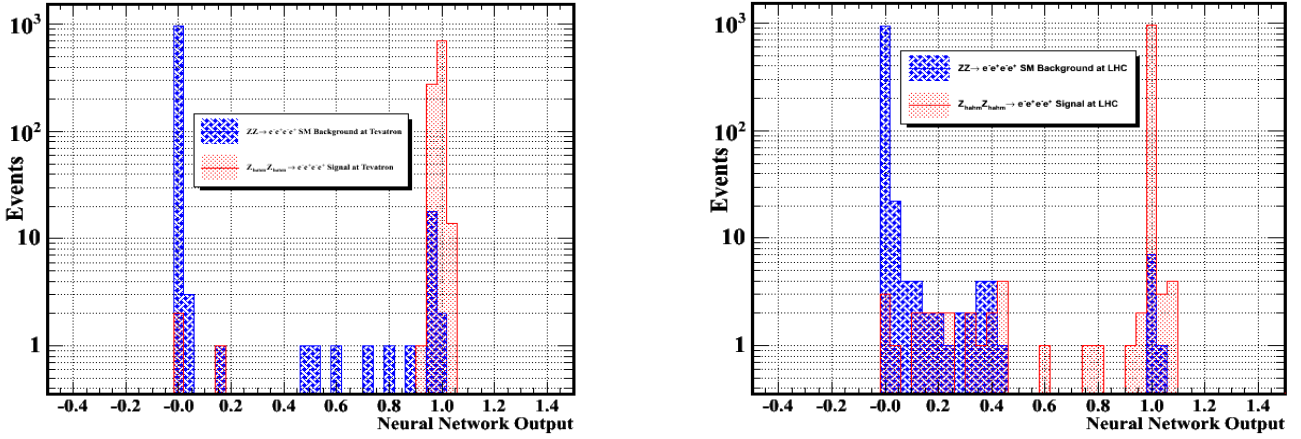

2-The Impact of variables influences on the network for $e^-e^+e^-e^+$ $final\ state$ where lInvm s the invariant mass, lPt is the transverse momentum and lEt is the transverse momentum

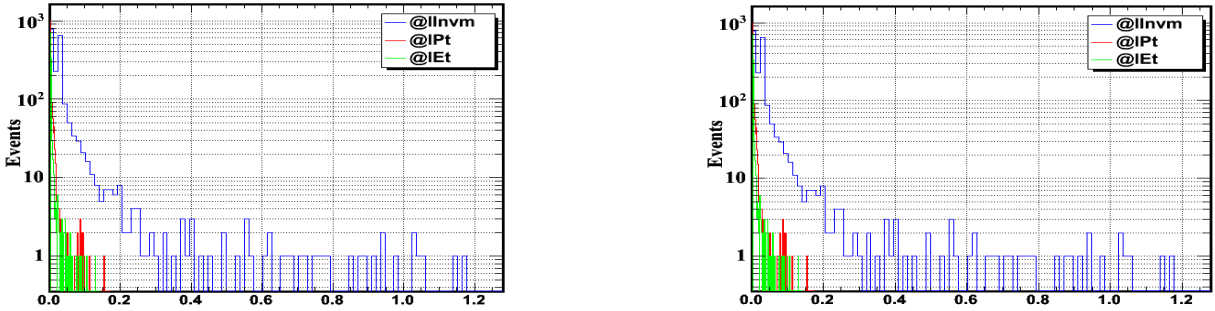

3-The error of the training and the test samples of the ANNs for $e^-e^+e^-e^+$ final state

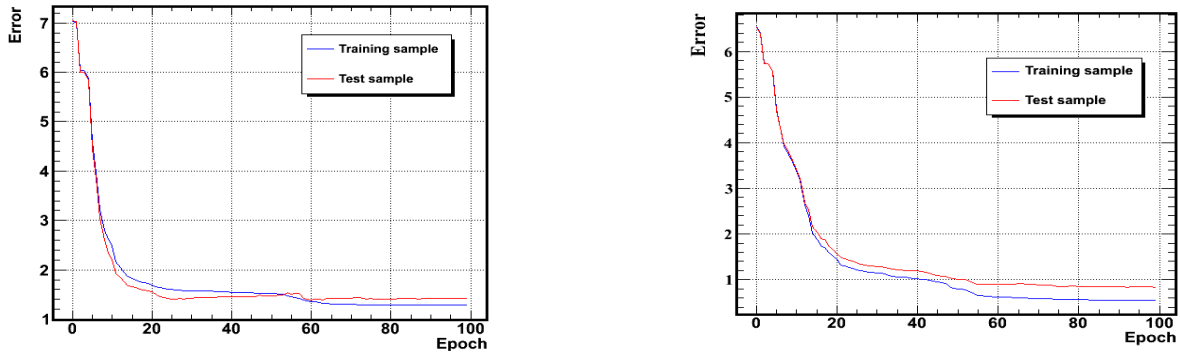

FIG. 9: Comparison between the ANNs Output discriminations of $e^-e^+e^-e^+$ final state produced from the decay of $Z_{HAHM}$ boson in Hidden Abelian Higgs Model (HAHM) at the LHC for $\sqrt{s} = 14\ TeV$ (LEFT) and at the Tevatron for $\sqrt{s} = 1.96\ TeV$ (RIGHT).

# 3 Scalar Sector of HAHM (New Higgs Boson H)

The signal detection of the new heavy Higgs boson decaying into a pair of the Standard Model Z bosons in the context of the HAHM is defined by the presence of two bottom jets from one Z boson decaying *hadronically* and the one Z boson decaying *leptonically* and the new Higgs final state is $\ell^\pm \ell^\mp jj$, $\ell = e, \mu$, or new Higgs boson decaying into a pair of Standard Model W bosons that decays leptonically with final state $\ell^\pm \ell^\mp \nu \bar{\nu}$, $\ell = e, \mu$ and missing energy. One the other hand, the new boson observed by the LHC detectors (ATLAS and CMS) [12, 13] is a Standard Model Higgs boson with mass 126 GeV. The experimental results on ratio of the observed cross section to the SM expectation in the four leptons channels is 1.4 from ATLAS experiment and is 0.7 from CMS experiment. The total cross sections for the Standard Model Higgs boson production at the LHC with mass of 125 GeV are predicted to be 17.5 pb for $\sqrt{s} = 7$ TeV and 22.3 pb for $\sqrt{s} = 8$ TeV [52, 53]. We used Monte Carlo programs to generate the scattering processes, patron showers and their hadronization, and to simulate the underlying event [14–16], Monte Carlo Pythia8 event generator [17] used to simulate the signal and the background processes. The HAHM new heavy Higgs boson production processes as in Figures 10, 11 considered in this work are the dominant gluon-gluon fusion gg → H and the signal cross section is computed at tree level using MadGraph5/MadEvent [18]. As we discussed, the final state consists of four charged leptons that produced from decaying a pair of new gauge boson $Z_{HAHM}$ in Hidden Abelian Higgs Model (HAHM) at the LHC for $\sqrt{s} = 10, 12$ and 14 TeV and at the Tevatron for $\sqrt{s} = 1.96$ TeV, also the production cross section is computed when new Higgs decaying to $Z_{HAHM}$ mixing with Z boson at the LHC and Tevatron.

The decay channels and branching ratios of the HAHM Higgs boson as a function of its mass are calculated using the CALCHEP as shown in Figure 12. The decay channel of the HAHM Higgs boson to two Standard Model Higgs has the highest ratio approx. 65% while the decay channel to pair of WW bosons is equal to 20% and decay to ZZ bosons is equal to 14%. Also, the new Higgs boson can decay hadronically with small ratio. In current work, we used the two decay channels H→ZZ→$\ell^\pm \ell^\mp jj$, $\ell = e, \mu$ and H→WW→ $\ell^\pm \ell^\mp \nu \bar{\nu}$, $\ell = e, \mu$ and missing energy to detect the signal of new Higgs boson at the LHC CERN (CMS and ATLAS detectors) and the Tevatron (CDF and D0 detectors). We generated events for two processes in the context of Hidden Abelian Higgs Model using MadGraph5/MadEvent. The selection events cuts for charged leptons $p_T > 50$ GeV and $|\eta| < 2.5$. We presented the analysis for different final states with electrons and muons as $(\bar{p}) \to H \to ZZ \to e^- e^+ b\bar{b}$, $pp(\bar{p}) \to H \to ZZ \to \mu^- \mu^+ b\bar{b}$, $pp(\bar{p}) \to H \to W^- W^+ \to e^- e^+ \bar{\nu}_e \nu_e$, $pp(\bar{p}) \to H \to W^- W^+ \to \mu^- e^+ \bar{\nu}_\mu \nu_e$ and $pp(\bar{p}) \rightarrow H \rightarrow W^-W^+ \rightarrow \mu^-\mu^+\bar{\nu}_\mu\nu_\mu$. We computed the different kinematics for all final states as invariant mass, transverse momentum, transverse energy and pseudorapidity where there is a peak at invariant mass value equal to 300 GeV of final state $e^-e^+b\bar{b}$ with jets that produced from decaying ZZ bosons at the LHC and the Tevatron this value of the invariant mass compared to the SM background value which equal to 240 as shown in Figure 14. One the other hand, in case of $\mu^-e^+\bar{\nu}_\mu\nu_e$ final state with missing energy that produced from decaying WW boson there is a peak at invariant mass value is 300 GeV compared with the SM background as in Figure 16. In Figure 15 we used the artificial neural networks for discrimination the signal and background events which is very efficient technique where we used three parameter as inputs for the neural network, invariant mass, transverse momentum and the transverse energy for each final state and then we will get the type of event signal or background as shown. We can get the signal of new Higgs boson at the hadrons colliders via signal of two leptons and jets or two leptons and missing energy.

## 3.1 Production of New Higgs Boson

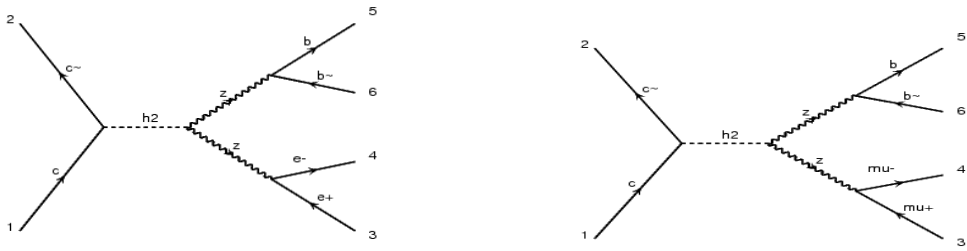

FIG. 10: Feynman diagrams of new Higgs boson h2 production and decay to two charged leptons and two jets.

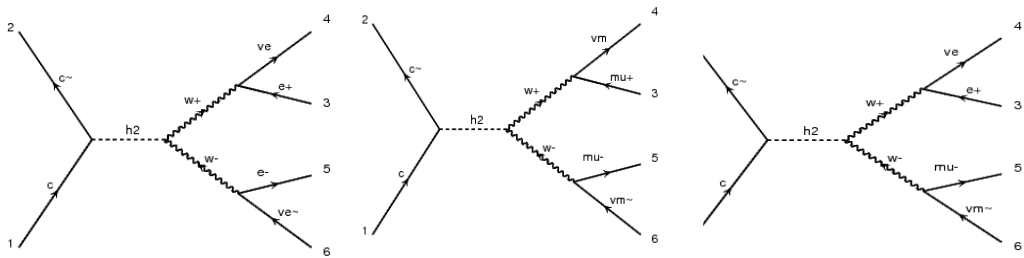

FIG. 11: Feynman diagrams of new Higgs boson h2 production and decay to two charged leptons and missing energy.

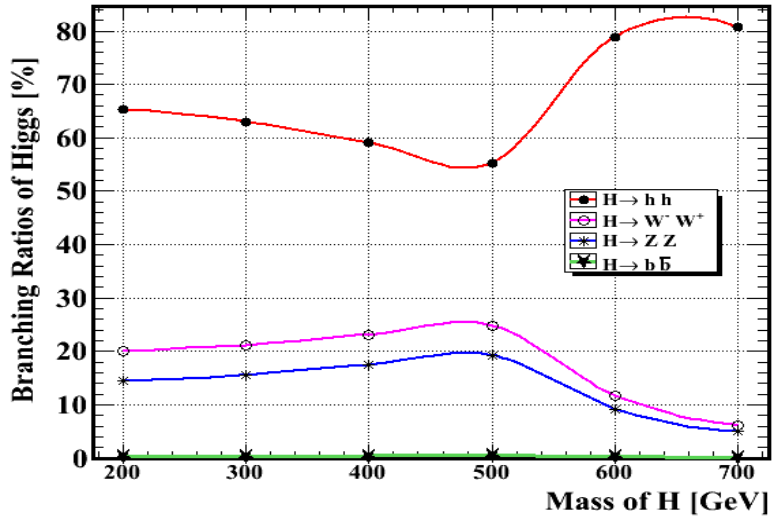

FIG. 12: All decay channels and branching ratios of Higgs boson as a function of the mass.

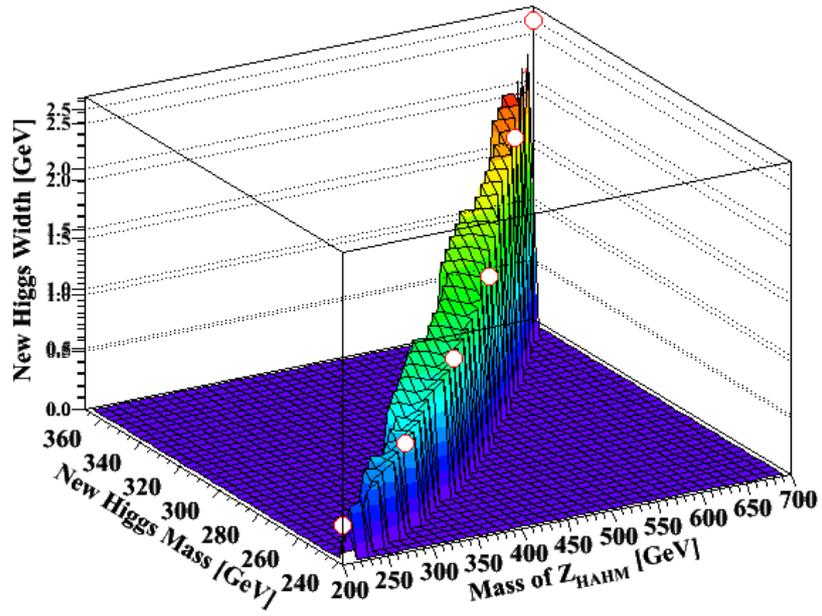

FIG. 13: Decay Width of the New Higgs boson as a function of its mass and $Z_{HAHM}$ boson mass

## 3.2 Detection Higgs Signal at Hadron Colliders

**3.2.1- Process** $pp(\bar{p}) \to H \to ZZ \to e^-e^+b\bar{b}$

**A-  Different distributions of ($e^-e^+b\bar{b}$) at the LHC (LEFT) and the Tevatron (RIGHT)**
1- Invariant Mass of $e^-e^+b\bar{b}$ $final\ state$ **[LHC (Left) & Tevatron (Right)]**

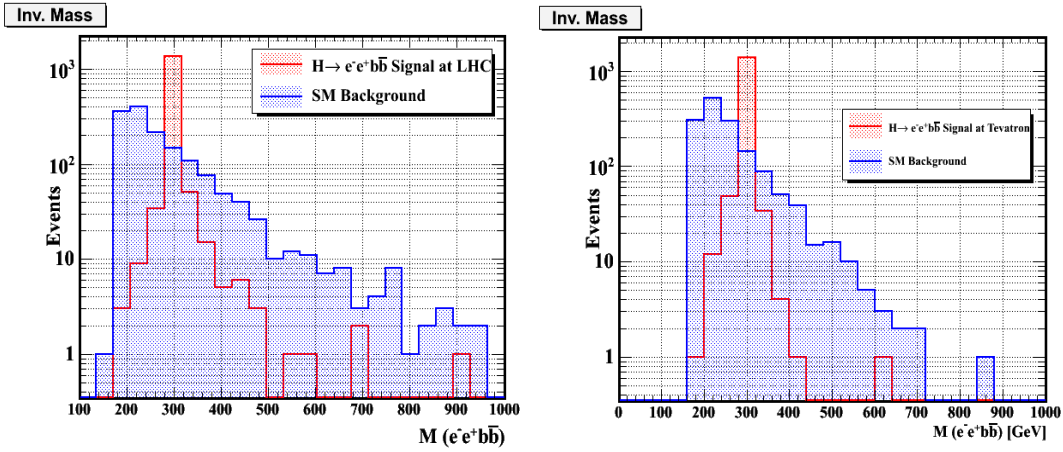

2- Transverse Momentum of $e^-e^+b\bar{b}$ $final\ state$

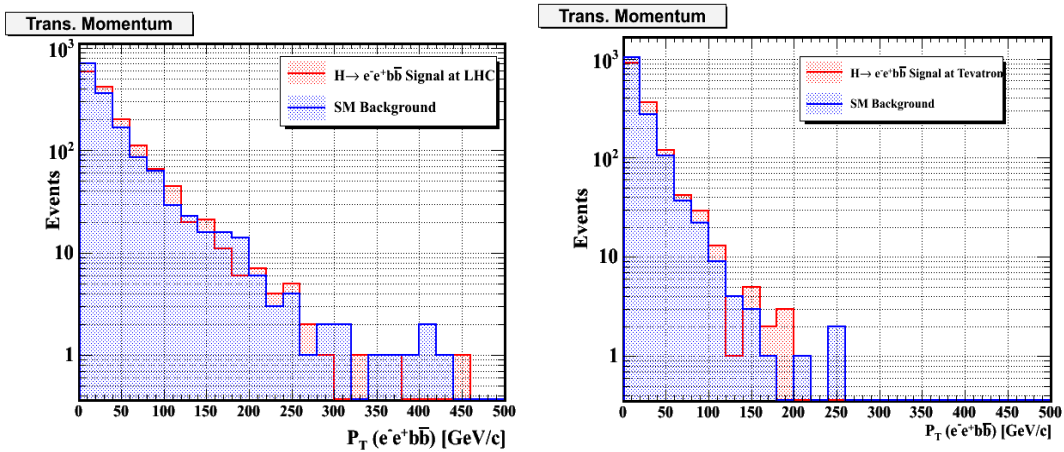

3- Transverse Energy of $e^-e^+b\bar{b}$ final state **[LHC (Left) & Tevatron (Right)]**

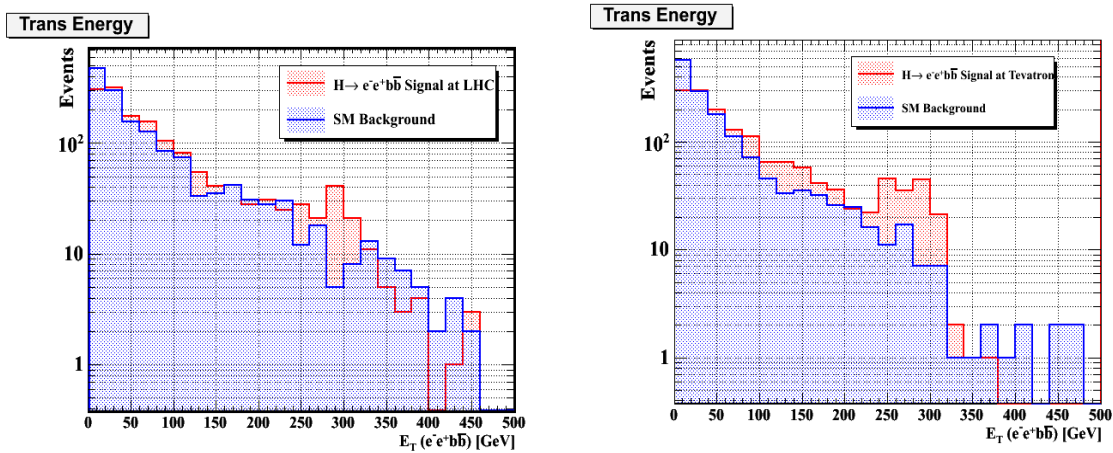

3- Pseudorapidity of $e^-e^+b\bar{b}$ $final$ $state$ **[LHC (Left) & Tevatron (Right)]**

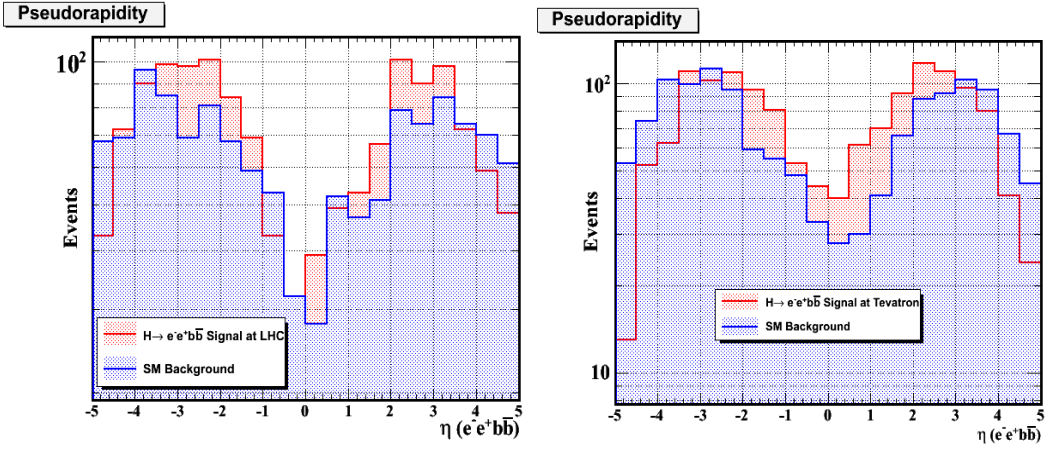

FIG. 14: Comparison between the different distributions of $e^-e^+b\bar{b}$ final state produced from the decay of new Higgs boson in Hidden Abelian Higgs Model (HAHM) at the LHC for $\sqrt{s} = 14\ TeV$ (LEFT) and at the Tevatron for $\sqrt{s} = 1.96\ TeV$ (RIGHT).

**B- The Discrimination between the New Higgs *Signal*s and *Backgrounds* events using Artificial Neural Networks (ANNs) at the *LHC* (LEFT) and the *Tevatron* (RIGHT)**

1- Final discriminations of ANNs Output (signal and background) for $e^-e^+b\bar{b}$ final state

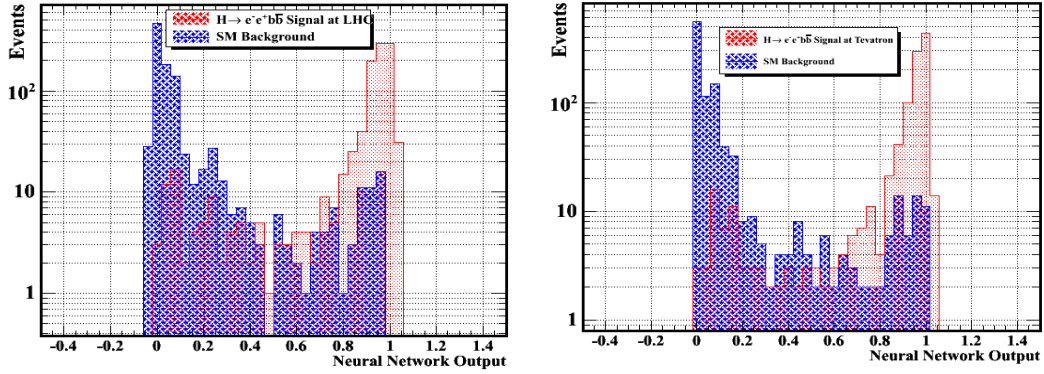

2-The Impact of variables influences on the network for e⁻e⁺b$\bar{b}$ $final$ $state$ where lInvm is the invariant mass, lPt is the transverse momentum and lEt is the transverse momentum.

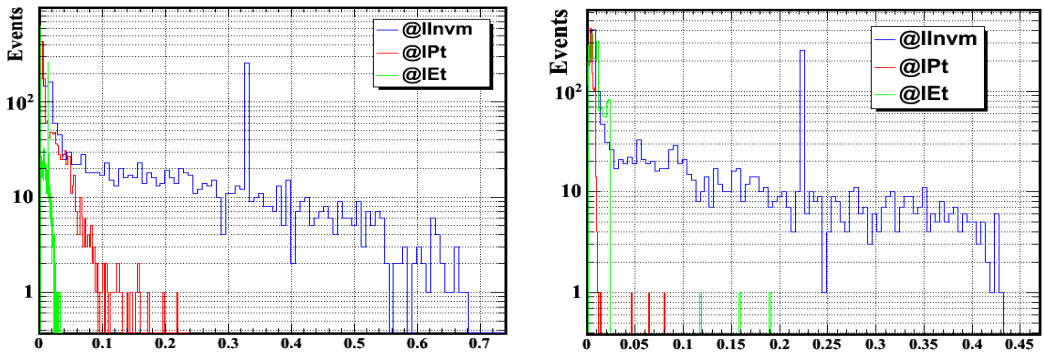

3-The error of the training and the test samples of the ANNs for $e^-e^+b\bar{b}$ final state

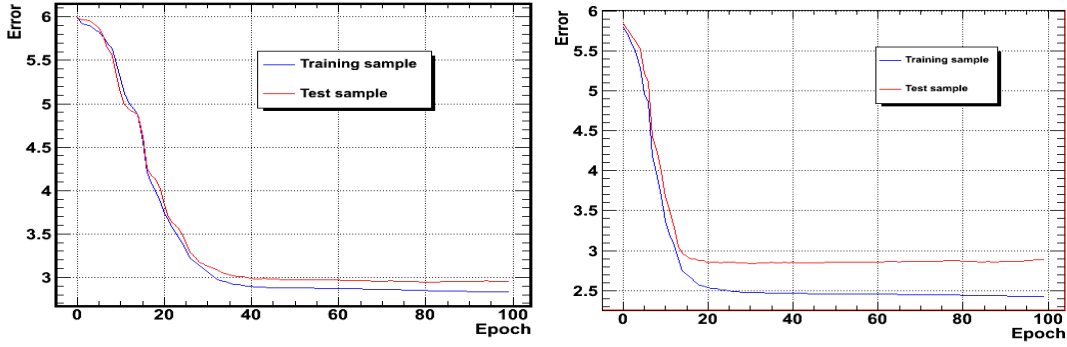

FIG. 18: Comparison between the ANNs Output discriminations of $e^-e^+b\bar{b}$ final state produced from the decay of new Higgs boson in Hidden Abelian Higgs Model (HAHM) at the LHC for $\sqrt{s} = 14\,TeV$ (LEFT) and at the Tevatron for $\sqrt{s} = 1.96\,TeV$ (RIGHT).

## 3.2.2 Process $pp(\bar{p}) \to H \to W^-W^+ \to \mu^-e^+\bar{\nu}_\mu\nu_e$

**A-Different distributions of ($\mu^-e^+\bar{\nu}_\mu\nu_e$) at the LHC (LEFT) and the Tevatron (RIGHT)**

1-Invariant Mass of $\mu^-e^+\bar{\nu}_\mu\nu_e$ $final\ state$ **[LHC (Left) & Tevatron (Right)]**

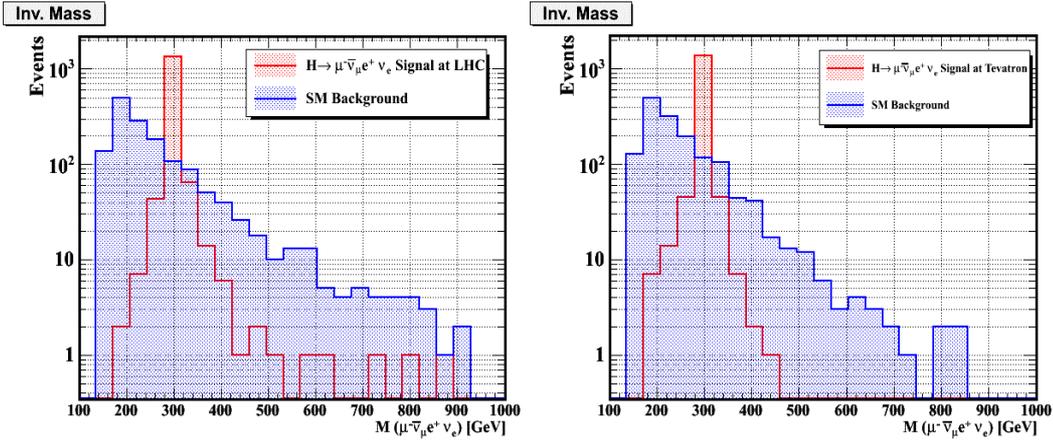

2-Transverse Momentum of $\mu^-e^+\bar{\nu}_\mu\nu_e$ $final\ state$ **[LHC (Left) & Tevatron (Right)]**

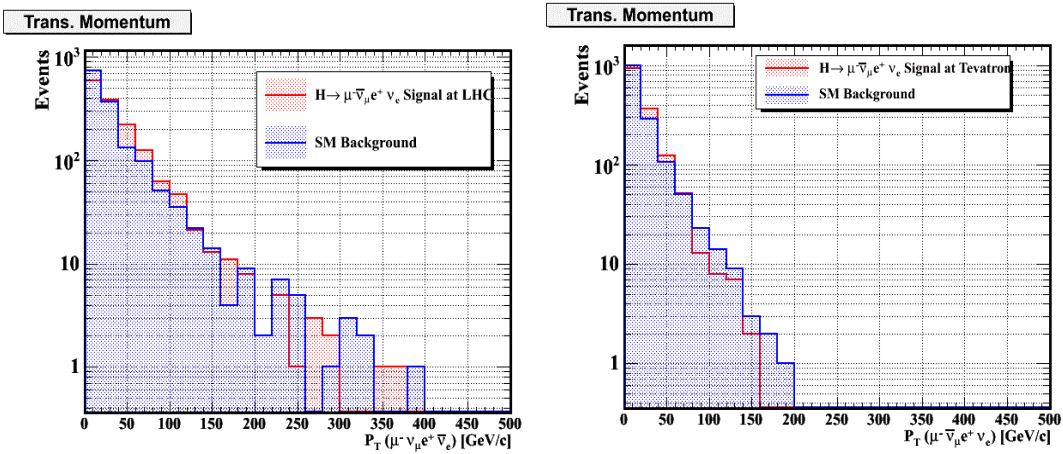

3- Transverse Energy of $\mu^- e^+ \bar{\nu}_\mu \nu_e$ final state [LHC (Left) & Tevatron (Right)]

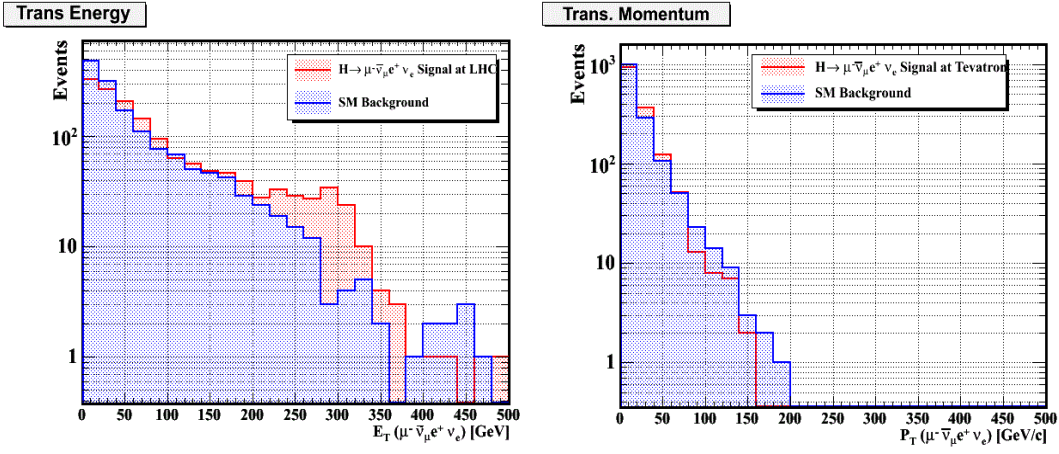

4- dorapidity of $\mu^- e^+ \bar{\nu}_\mu \nu_e\ final\ state$ [LHC (Left) & Tevatron (Right)]

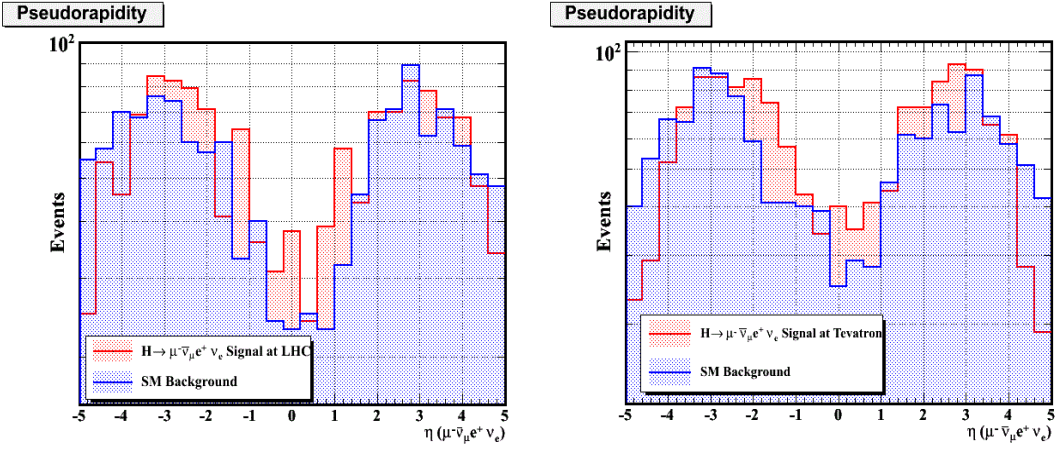

FIG. 16: Comparison between the different distributions of $\mu^- e^+ \bar{\nu}_\mu \nu_e$ final state produced from the decay of new Higgs boson in Hidden Abelian Higgs Model (HAHM) at the LHC for $\sqrt{s} = 14\ TeV$ (LEFT) and at the Tevatron for $\sqrt{s} = 1.96\ TeV$ (RIGHT).

**B-The Discrimination between the New Higgs *Signal*s and *Backgrounds* events using Artificial Neural Networks (ANNs) at the *LHC* (LEFT) and the *Tevatron* (RIGHT)**

1- Final discriminations of ANNs Output (signal and background) for $\mu^- e^+ \bar{\nu}_\mu \nu_e$ final state

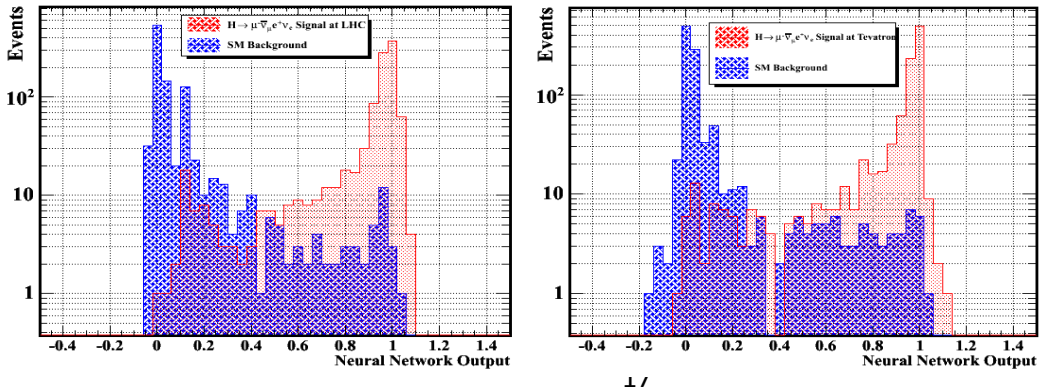

2-The Impact of variables influences on the network for $\mu^- e^+ \bar{\nu}_\mu \nu_e$ $final\ state$ where lInvm is the invariant mass, lPt is transverse momentum and lEt is the transverse momentum.

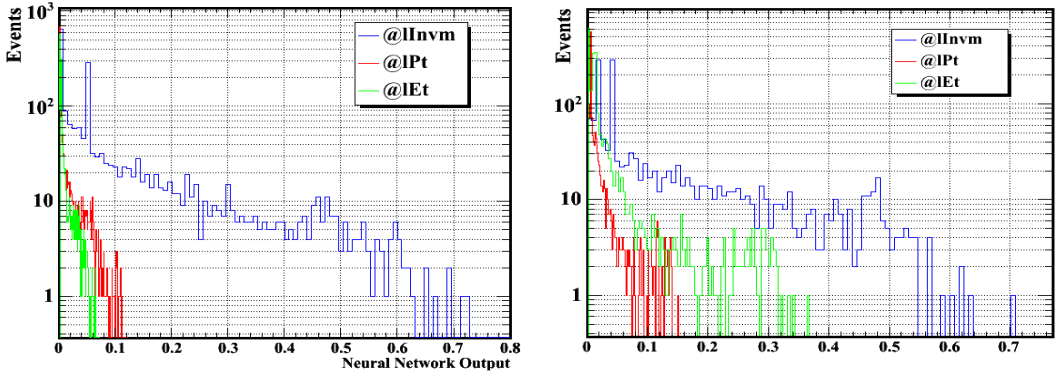

3-The error of the training and the test samples of the ANNs for $\mu^- e^+ \bar{\nu}_\mu \nu_e$ final state

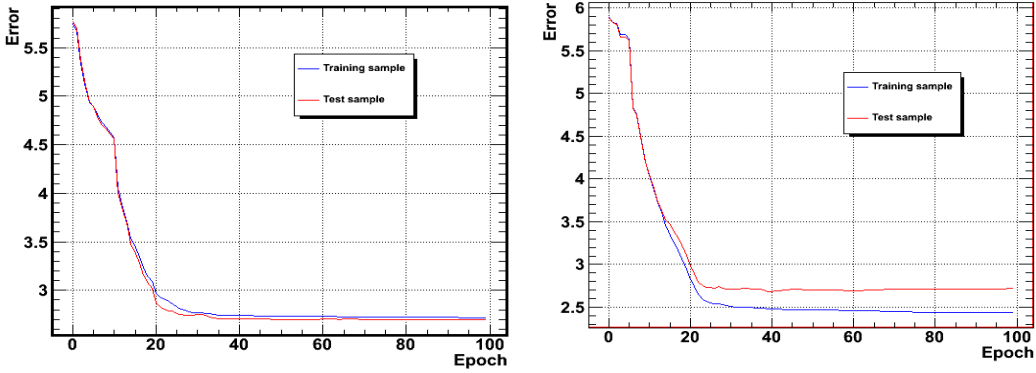

FIG. 17: Comparison between the ANNs Output discriminations of $\mu^- e^+ \bar{\nu}_\mu \nu_e$ final state produced from the decay of new Higgs boson in Hidden Abelian Higgs Model (HAHM) at the LHC for $\sqrt{s} = 14\ TeV$ (LEFT) and at the Tevatron for $\sqrt{s} = 1.96\ TeV$ (RIGHT).

## 4 Conclusions

In the context of the Hidden Abelian Higgs Model (HAHM), we have performed two searches using Monte Carlo data at the LHC and at the Tevatron. The first search is for new neutral massive gauge boson $Z_{HAHM}$ with 4 charged leptons in the final state in the mass range $200\ \text{GeV} \leq Z_{HAHM} \leq 700\ \text{GeV}$. From our simulation for $Z_{HAHM}$ pair production, the best decay channel search for $Z_{HAHM}$ boson is $Br(Z_{HAHM} \to \ell^\pm \ell^\mp) = 13\%$ and from the reconstruction mass of $Z_{HAHM}$ pair form four charged leptons invariant masses, the $Z_{HAHM}$ pair production may be produced at 500 GeV at the LHC and 450 GeV at the Tevatron. The second search is a new Higgs boson using two final states. The first final state includes 2 charged leptons plus jets and the second final state has 2 charged leptons plus missing energy. From our simulation for new Higgs production, there are two decay channels which are the best to search for new Higgs, $Br(H \to W^- W^+) = 20\%$ with final state $\ell^\pm \ell^\mp \bar{\nu}\nu$ and $Br(H \to ZZ) = 15\%$ with final state $\ell^{\pm}\ell^{\mp}b\bar{b}$ ($\ell = e, \mu$). We reconstructed the new Higgs mass with invariant masses from the two final states. It is found that the new Higgs may be found at the LHC and the Tevatron with mass 300 GeV

## AKNOWLEDGMENT

It is a pleasure to thank Prof. Torbjorn Sjostrand, Department of Theoretical Physics, Lund University, Lund, Sweden, the main author of Monte Carlo Event Generator (MCEG) Pythia8 for useful discussions. The work by Maxim Yu. Khlopov was supported by the Ministry of Education and Science of Russian Federation project 3.472.2014/K and grant RFBR 14-22-03048.